\documentclass[aps,prl,twocolumn,preprintnumbers,
superscriptaddress,floatfix]{revtex4}

\usepackage{amssymb,multirow}
\usepackage{graphicx}

\begin{document}

\newcommand{\Tr}{\mbox{Tr\,}}
\newcommand{\beq}{\begin{equation}}
\newcommand{\eeq}{\end{equation}}
\newcommand{\bea}{\begin{eqnarray}}
\newcommand{\eea}{\end{eqnarray}}
\renewcommand{\Re}{\mbox{Re}\,}
\renewcommand{\Im}{\mbox{Im}\,}

\voffset 1cm

\newcommand\sect[1]{\emph{#1}---}

\pagestyle{empty}

\preprint{
\begin{minipage}[t]{3in}
\begin{flushright} NI-08-006
\\
SHEP-08-07
\\[30pt]
\hphantom{.}
\end{flushright}
\end{minipage}
}

\title{Mesonic quasinormal modes of the Sakai-Sugimoto model at high
temperature}

\author{Nick Evans$^{1,2}$ and Ed Threlfall}

\affiliation{School of Physics and Astronomy, University of
Southampton,
Southampton, SO17 1BJ, UK \\
$^2$Newton Institute for Mathematical Sciences,
20 Clarkson Road, Cambridge, CB3 0EH, UK\\
evans@phys.soton.ac.uk,ejt@phys.soton.ac.uk}

\begin{abstract}
\noindent We examine the mesonic thermal spectrum
of the Sakai-Sugimoto  model of holographic QCD by finding the
quasinormal frequencies of the supergravity dual.  If flavour is
added using D8-$\bar{D8}$ branes there exist embeddings where the
D-brane worldvolume contains a black hole.  For these embeddings
(the high-temperature phase of the Sakai-Sugimoto model) we
determine the quasinormal spectra of scalar and vector mesons
arising from the worldvolume DBI action of the D-brane. We stress
the importance of a coordinate change that makes the in-falling
quasinormal modes regular at the horizon allowing a simple
numerical shooting technique. Finally we examine the effect of
finite spatial momentum on quasinormal spectra.
\end{abstract}

\maketitle

\section{Introduction}

Gravity dual descriptions \cite{Malda,Witten:1998qj,Gubser:1998bc}
of strongly coupled gauge theories with quarks
have recently shed light on the physics of mesons and chiral symmetry
breaking \cite{Erdmenger:2007cm}.
There has also been considerable interest in studying the finite temperature
behaviour of these systems.

The simplest such dual is the near horizon geometry of a D3-D7 brane system
which describes an ${\cal N}=2$ gauge theory. The D7 branes can be treated
as probes \cite{Karch} in the limit where the number of flavours is much less
than the number of colours, $N_f \ll N_c$, or the full back reacted
geometry can be found \cite{Polchinski,Bertolini:2001qa, Kirsch}.
The meson spectrum in the probe limit has been
computed in \cite{Mateos}. Finite temperature manifests itself as the presence
of a black hole in the dual space-time \cite{Witten:1998qj}.
In the infinite volume limit the
black hole geometry is energetically preferred for any temperature greater
than zero. The transition from an AdS space corresponds to the analogue of the
deconfinement transition in the pure glue gauge theory
(which in a conformal theory
occurs as soon as the dimensionful parameter, T, is introduced). A further
first order phase transition has also been found in this system
\cite{Babington,Apreda,Mateos1,Mateos2,Albash}
when the
temperature passes through the scale of the mass of the mesonic bound states
- this corresponds to when the horizon of the black hole grows to
swallow the D7 probe in the interior of the space. This transition
has an associated small jump in the chiral condensate's value but the
main physics of the transition appears to be the meson fields melting
into the thermal background. Once the D7 brane enters the horizon there
are no longer normalizable fuctuations of the D7 brane that generate a
discrete set of meson bound states. Instead there are quasi-normal modes
of the black hole corresponding to fluctuations of the D7 brane that
are pure in-falling at the horizon. These fluctuations correspond in the
dual gauge theory to excitations of the plasma with a complex mass parameter -
the excitations have both a mass and a decay time. The spectrum of these
quasi-normal modes has been explicitly computed in \cite{Hoyos:2006z}.

Another interesting model is the Sakai-Sugimoto model
\cite{Sakai:2004cn,Sakai:2005yt} which is based on a
(wrapped) D4 D8 $\bar{D8}$ system. It
is a gravity dual of a non-supersymmetric
gauge theory (which is four dimensional in the IR but five dimensional
in the UV) that dynamically breaks a non-abelian chiral symmetry
of its quark fields. The high temperature phase again corresponds to a
transition to a black hole geometry. The transition occurs when the black
hole's radius becomes of order the wrapped circumference of the D4 brane
which is also the parameter that determines the mass gap of the theory.
This behaviour is more akin to what one would expect in QCD than that of
the conformal theory discussed above.

Massless chiral quarks can be introduced by placing the probe D8 and $\bar{D8}$
branes at anti-podal points on the circle the D4 brane is wrapped on.
In the near horizon limit of the D4 branes these D8 branes choose to join at
the scale of the mass gap breaking the chiral symmetries on their world
volumes to the diagonal sub-group and generating a mass gap for the
mesonic fluctuations of the D8s. When the geometry makes the transition at
finite temperature to the black hole background the D8 and $\bar{D8}$
disconnect and instead lie straight and fall into the horizon
\cite{Aharony:2006y}.
Chiral symmetry breaking is therefore restored along with deconfinement.

There is a larger class of embeddings in which the D8 and $\bar{D8}$ join
at a larger radius in the space so there is a bigger mass gap for the quarks.
In \cite{EvansThrelfall:2007a} we have argued that these embeddings
describe a quark mass in the
theory although it has been also argued in the literature \cite{Antonyan}
that the chiral
symmetry breaking scale is being enhanced in these cases by higher dimension
operators. The distinction is not important for what we discuss here - in these
cases there is a further first order transition as the temperature
(horizon) grows through the mass scale of the mesons. The transition is very
much like that of the D3-D7 system in that the mesonic fluctuations of the
D8 branes are replaced by quasinormal modes of the black hole. The mesons
of the theory have melted into the plasma.

The mesonic fluctuations of the D8 branes above the
phase transition have been studied in \cite{Zamaklar}. Here we will concentrate on the
very high temperature phase where the D8 branes lie straight and fall into
the black hole horizon. We will explicitly compute the quasinormal mode
spectrum corresponding to the scalar and vector mesons of the
theory.

As a prelude to this we compute the quasinormal spectrum of a
Klein-Gordon scalar living on the D8 brane worldvolume.  We
apply the idea of regularizing the coordinates for ingoing modes,
which has previously been used in asymptotically-flat spacetimes
(see for example \cite{Dolan:2005a}) and in the context of AdS-CFT
in \cite{Horowitz}.  The result of this is that the ingoing mode
is described by a regular Taylor series at the black hole horizon.
We use this as the initial condition and obtain the quasinormal
spectra by shooting out from the horizon. We wish to stress that
this is a much cleaner numerical process than trying to match on
to oscillating solutions at the horizon.  We use the same method
to examine the spectra of modes arising from the DBI action of an
embedded D8 brane.  We treat a scalar fluctuation of the brane in
the geometry and a Lorentz vector arising from the Maxwell field
on the D-brane. Finally we briefly discuss the effect of nonzero
momentum on the spectra and extract the diffusion coefficient from
the lowest quasinormal modes of longitudinal vector excitations
in the small $k$ `hydrodynamic' limit.
\bigskip

\section{The Geometry}

The metric of the high temperature Sakai-Sugimoto model is
\begin{equation} \begin{array}{ccc}
ds^2&=&\left ( \frac{u}{R} \right )^{\frac{3}{2}} \left ( -f(u) dt^2+dx_3^2
+d \tau^2 \right ) \\&&\\
&& + \left ( \frac{R}{u} \right )^{\frac{3}{2}} \left(
\frac{du^2}{f(u)}+u^2 d\Omega_4^2 \right) \end{array}
\end{equation}
Here $f(u)=1-\left ( \frac{u_T}{u}\right )^3$ and the dilaton is $e^{-\phi}
= g_s^{-1} \left ( \frac{R}{u} \right )^{\frac{3}{4}}$.

The parameter $u_T$, representing the position of the horizon in the
geometry, gives the temperature in the dual field theory by the relation
$T=\frac{3}{4 \pi} \frac{u_T^{\frac{1}{2}}}{R^{\frac{3}{2}}}$.  This is the
Hawking temperature of the black hole.

Let us work in the dimensionless radial coordinate $x \equiv
\frac{u}{u_T}$ and measure the Minkowski and $\tau$ dimensions in
units of $\sqrt{\frac{R^3}{u_T}}$.   This corresponds to measuring
frequencies and momenta in units $\propto T$.

The metric becomes
\begin{equation} \label{metric} \begin{array}{ccc}
\frac{ds^2}{R^{\frac{3}{2}} \sqrt{u_T}} &=& x^{\frac{3}{2}} \left (-f(x) dt^2
+ dx_3^2 + d\tau^2 \right ) \\&\\
&& + x^{-\frac{3}{2}} \left ( \frac{dx^2}{f(x)} + x^2 d\Omega_4^2 \right )
\end{array}
\end{equation}

We will consider the spectrum of modes associated to a D8 brane.
The D8 branes fill the space except for the $\tau$ direction in
which they live at a single value of $\tau = \tau_0$ - this is the
energetically preferred high temperature configuration
\cite{Aharony:2006y}.  The induced metric on the D8 brane
worldvolume is just (\ref{metric}) with $d\tau=0$.

\section{Regular horizon coordinates for quasinormal modes}

As a warm up we shall first consider linear fluctuations of a
Klein-Gordon scalar restricted to the worldvolume of the D8
brane.  This is not a physical mesonic state of the field theory
but it exhibits similar quasinormal modes in the supergravity dual
and we use it to illustrate our calculational technique. The 9D action describing the fluctuation is
\begin{equation}
\mathcal{S}_{SF}= \frac{1}{2} \int d^{9}x 
\sqrt{-g} g^{ab} \nabla_a \Phi \nabla_b \Phi
\end{equation}
Here the geometry is given by the induced metric on the D8 brane.

The equation of motion for the fluctuation is
\begin{equation}
\frac{1}{\sqrt{-g}} \partial_a \left ( \sqrt{-g} g^{ab}
\partial_b \Phi \right )=0
\end{equation}
Writing our the equation for a scalar fluctuation with spatial momentum
$k$ with respect to the plasma rest frame and zero $S_4$ spin as
$\Phi \propto e^{- i \omega t+i k \cdot x_3}$ one
obtains the equation
\begin{equation} \label{glueeom}
\left ( x^{\frac{19}{4}} f(x) \Phi' \right )' + x^{\frac{7}{4}} \left ( \frac{\omega^2}{f(x)}-k^2 \right ) \Phi=0
\end{equation}
Here the prime indicates an $x$ derivative.
The large-$x$ asymptotic of the equation is $\left ( x^{\frac{19}{4}} \Phi' \right )'=0$
with solution $\Phi \sim c_1+c_2 \; x^{-\frac{15}{4}}$.
For a normalizable solution we clearly want the decaying power.

\begin{centering}
\begin{figure*}
\begin{centering} \begin{tabular}{cc}
\includegraphics[width=80mm]{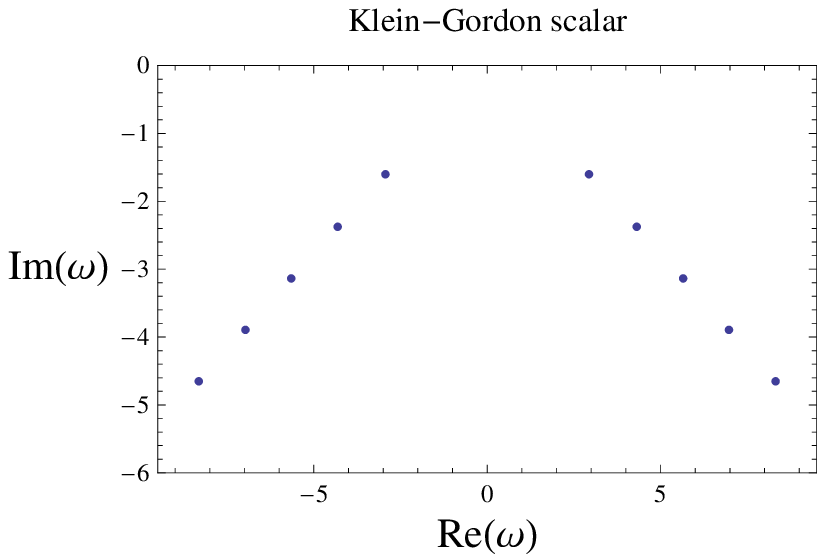} &
$\left. \right.$ \hspace{7cm} \end{tabular} \vspace{-4.5cm}

\begin{tabular}{cc}
$\left. \right.$ \hspace{7cm} &
\begin{tabular}[t]{|c|c|}
\hline
$n$&$\Omega_n$\\
\hline
1& $\pm$ 2.9348 - 1.6032 $i$\\
2& $\pm$ 4.3133 - 2.3760 $i$\\
3& $\pm$ 5.6509 - 3.1367 $i$\\
4& $\pm$ 6.9725 - 3.8928 $i$\\
5& $\pm$ 8.3151 - 4.6531 $i$\\
\hline
\end{tabular} \end{tabular} \vspace{1cm}

\par
\end{centering}
\caption{\small{The lowest five $k=0$ Klein-Gordon scalar quasinormal
frequencies in the complex $\omega$ plane.}}
\end{figure*}
\par
\end{centering}

Taking the near horizon limit ($x \rightarrow 1$)
one finds $\Phi \sim (x-1)^{\pm i
\frac{\omega}{3}}$.  Since the whole solution is $\exp(-i\omega t
\pm i \frac{\omega}{3} \ln(x-1))$ the solutions are ingoing ($-$)
and outcoming ($+$) waves which oscillate infinitely many times
before reaching the horizon.

In these coordinates it will be hard to find the full solution numerically -
one needs to shoot out from, or on to, a highly oscillatory solution near
the horizon. A better way to proceed is to change coordinates to make
the infalling solution regular at the horizon so numerical methods can
be more easily used. In particular we will shift coordinates so
\beq t = h - \alpha(x) \eeq
To make the infalling solution regular we require
\begin{equation}
\alpha(x) =
\frac{1}{3} \ln(x-1) + \; ... \;
\end{equation}
where the additional terms are regular at the horizon.  One way of satisfying this is to define
\begin{equation}
\frac{\partial \alpha}{\partial x}= \frac{1}{x^3-1}
\end{equation}
We then have
\begin{equation}
dh=dt+\frac{1}{x^3 f(x)} \; dx
\end{equation}
leaving the metric in the new coordinates as
\begin{equation} \begin{array}{ccc}
ds^2&=&-x^{\frac{3}{2}} f(x) dh^2+2 x^{-\frac{3}{2}} dh \; dx+x^{-\frac{3}{2}} dx^2 \\ &&\\
&& +x^{\frac{3}{2}}
\left ( dx_3^2+d \tau^2\right )+ \sqrt{x} d \Omega_4^2 \end{array}
\end{equation}
The induced metric is the above with $d\tau=0$.
Note that the near horizon solution of (\ref{glueeom}) only depended
on the powers of $f$ in the function not the powers of $x$ (which
becomes one in the near horizon limit). These powers of $f$ will
turn out to be the same for all of the modes we consider below and
so this change of coordinates will suffice to make all infalling
modes we look at regular.

We can now recompute the scalar equation of motion and we find
\begin{equation}
\left ( x^{\frac{19}{4}} f \Phi' \right )'- i \omega \left (
2 x^{\frac{7}{4}} \Phi'+\frac{7}{4} x^{\frac{3}{4}} \Phi
\right ) + \left ( \omega^2-k^2 \right ) x^{\frac{7}{4}} \Phi=0
\end{equation}
We note that this is an equation with five singular points
in the complex $x$-plane. The general solution of such equations
is not given in terms of well-known functions, nor  can one
immediately apply the continued fraction method as used in
\cite{Leaver}.  Accordingly we use a purely numerical approach.

There are two criteria for a good solution.  Firstly the solution
should be a purely ingoing wave at the black hole horizon since
classically a black hole can absorb but not emit particles.
Secondly the solution must be normalizable when integrated over
the D-brane worldvolume.

The large-$x$ asymptotic of the equation is the same as we found
in the usual `Schwarzschild' coordinates and so we choose the
decaying power which is normalizable as $x \rightarrow \infty$.

At the horizon we will seek a solution in the form of a
Frobenius series $\Phi=\Sigma_{n=0}^{\infty} a_n z^{n+s}$
in $z \equiv x-1$.  Now substituting this into the differential
equation yields the indicial equation

\begin{equation}
s \left ( s- \frac{2}{3}i \omega \right ) =0
\end{equation}

The solution with $s=0$ is regular at the horizon and corresponds
to a purely infalling solution.

Using the regular Taylor series for the infalling solution as the initial
condition we shoot out from the horizon.  By requiring our solution to vanish
as $x \rightarrow \infty$ we can find the quasinormal frequencies.
They are displayed in Fig.1.

\begin{centering}
\begin{figure*}
\begin{centering} \begin{tabular}{cc}
\includegraphics[width=80mm]{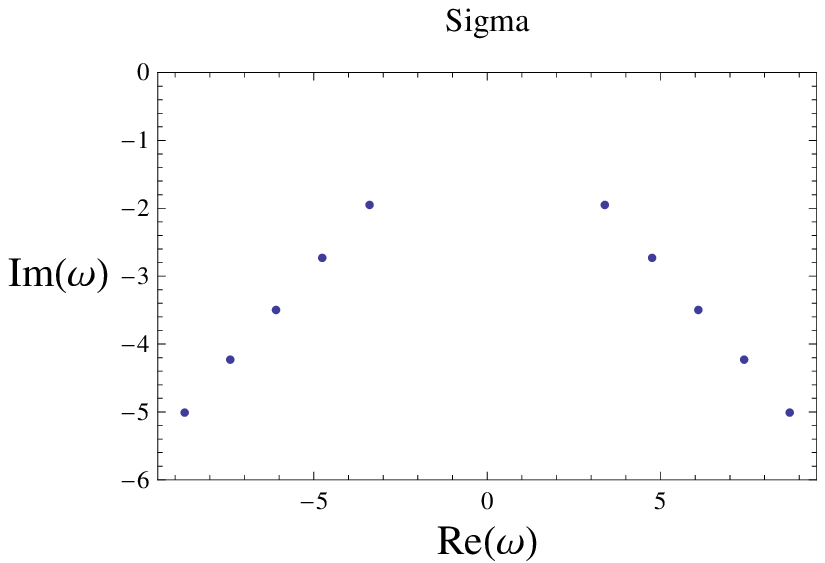} &
$\left. \right.$ \hspace{7cm} \end{tabular} \vspace{-4.5cm}

\begin{tabular}{cc}
$\left. \right.$ \hspace{7cm} &
\begin{tabular}[t]{|c|c|}
\hline
$n$&$\Omega_n$\\
\hline
1& $\pm$ 3.3896 - 1.9498 $i$\\
2& $\pm$ 4.7560 - 2.7320 $i$\\
3& $\pm$ 6.0900 - 3.4970 $i$\\
4& $\pm$ 7.4110 - 4.2297 $i$\\
5& $\pm$ 8.7234 - 5.0105 $i$\\
\hline
\end{tabular} \end{tabular} \vspace{1cm}

\par
\end{centering}
\caption{\small{The lowest five $k=0$ sigma meson quasinormal frequencies in the complex $\omega$ plane.}}
\end{figure*}
\par
\end{centering}

\section{Quasinormal modes from flavour branes}

We will now consider the physical quasinormal modes of D8 branes, coming from the DBI action.

The presence of the brane corresponds to the inclusion of a chiral
quark field in the gauge theory. Anomaly cancellation requires quarks to
occur in vector like pairs so there must naturally be a partner $\bar{D8}$.
The quasi-normal mode spectrum we compute below will therefore be parity
doubled in the gauge theory.

We will again use the coordinates (10) to make infalling solutions regular at the horizon.

\subsection{Scalar mesons}

We first analyze the scalar mode corresponding to a geometric fluctuation of the D8 embedding.
The DBI Lagrangian for this is
\begin{equation}
\mathcal{L} =e^{-\phi} \sqrt{-Det \left ( g_{MN} \frac{\partial x^M}
{\partial \xi^a} \frac{\partial x^N}{\partial \xi^b} \right )}
\end{equation}
 We will parameterize the fluctuation as $\tau=\tau_0+\phi(x)e^{-i \omega h+i k \cdot x_3}$
representing a mesonic excitation with spatial momentum $k$ relative to the plasma rest frame
with zero $S_4$ spin.  The Lagrangian is (a dot indicates an
$h$-derivative and a prime an $x$-derivative)
\begin{equation}
\mathcal{L}=  x^{\frac{5}{2}} \sqrt{1+
x^3 f(x) \phi'^2+2  \dot{\phi} \phi'-\dot{\phi}^2}
\end{equation}
Expanding the square root to quadratic order the equation
of motion is
\begin{equation}
\left ( x^{\frac{11}{2}} f \phi' \right )' - i \omega
\left (2 x^{\frac{5}{2}} \phi'+\frac{5}{2} x^{\frac{3}{2}} \phi \right ) +  \left( \omega^2-k^2 \right )
x^{\frac{5}{2}} \phi =0
\end{equation}
Using the regular Taylor series as initial condition we shoot out from the
horizon and requiring our solution to vanish as $x \rightarrow \infty$ we can
find the quasinormal frequencies. They are shown in Fig.2.

\begin{centering}
\begin{figure*}
\begin{centering} \begin{tabular}{cc}
\includegraphics[width=80mm]{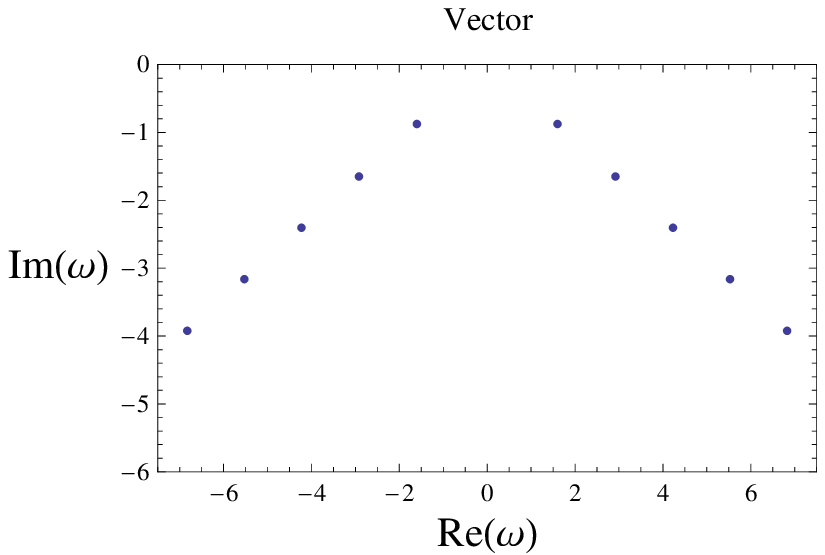} &
$\left. \right.$ \hspace{7cm} \end{tabular} \vspace{-4.5cm}

\begin{tabular}{cc}
$\left. \right.$ \hspace{7cm} &
\begin{tabular}[t]{|c|c|}
\hline
$n$&$\Omega_n$\\
\hline
1& $\pm$ 1.6016 - 0.8957 $i$\\
2& $\pm$ 2.9217 - 1.6512 $i$\\
3& $\pm$ 4.2265 - 2.4054 $i$\\
4& $\pm$ 5.5269 - 3.1611 $i$\\
5& $\pm$ 6.8264 - 3.9217 $i$\\
\hline
\end{tabular} \end{tabular} \vspace{1cm}

\par
\end{centering}
\caption{\small{The lowest five static vector meson quasinormal frequencies in the complex $\omega$ plane.}}
\end{figure*}
\par
\end{centering}

\subsection{Vector mesons - transverse}

We obtain the quasinormal spectrum for a Maxwell field on the D8 brane worldvolume,
which is dual to the quasinormal spectrum of vector mesons.  We use the ansatz
$A_{\mu}= \xi_{\mu} A(x) e^{i k \cdot x_4}$ which is a Lorentz vector with zero
$S_4$ spin.  The equation of motion is
\begin{equation}
\partial_a \left ( e^{-\phi} \sqrt{-g} F^{ab} \right )=0
\end{equation}
%

Fixing the gauge $k^{\mu}\xi_{\mu}=0$ one obtains the equation of motion
for a transversely-polarized Lorentz vector.
With this ansatz the only nontrivial equation is (for example choosing $k_2$ nonzero and the vector in the $1$-direction)
for $b=1$, ie $\partial_a (e^{-\phi} \sqrt{-g} F^{a1})=0$, giving
\begin{equation}
(x^{\frac{5}{2}}f A')'-i \omega \left ( 2 x^{-\frac{1}{2}} A'-\frac{1}{2} x^{-\frac{3}{2}}
A \right )+ \left (\omega^2-k^2 \right ) x^{-\frac{1}{2}} A=0
\end{equation}
Using the regular Taylor series as initial condition we shoot out from
the horizon and requiring our solution to vanish as $x \rightarrow \infty$
we can find the quasinormal frequencies. They are displayed in Fig.3.

\subsection{Vector mesons - longitudinal}

In the zero-temperature case the transverse mesons exhaust the vector spectrum.
For finite temperature however one can also identify a purely electric
longitudinal solution of the supergravity dual Maxwell field.
The longitudinal electric field we deal with is $E_1=k A_0+\omega A_1$
($k$ points along the $1$-axis making the vector potential curl-free).

There are three relevant equations of motion: first $\partial_a \left (e^{-\phi} \sqrt{-g} F^{ax} \right )=0$.
Writing this out for our choice of metric one obtains
\begin{equation}
i \omega A_0'+i k g^{11}g^{xx} A_1'=-g^{11} g^{x0} \left ( k^2 A_0+ \omega k A_1 \right )
\end{equation}
The second equation is $\partial_a \left (e^{-\phi} \sqrt{-g} F^{a0} \right )=0$.
Writing this out for our choice of metric one obtains
\begin{equation} \begin{array}{ccc}
0& =& -\left ( e^{-\phi} \sqrt{-g} A_0' \right )' -i k e^{-\phi}\sqrt{-g} g^{11}g^{0x} A_1'\\
&&-e^{-\phi} \sqrt{-g} g^{11}g^{00}
\left (k^2 A_0+k \omega A_1 \right )\end{array}
\end{equation}
Finally from $\partial_a \left (e^{-\phi} \sqrt{-g} F^{a1} \right )=0$ one obtains
\begin{equation} \begin{array}{ccc}
0 & = & \left (e^{-\phi} \sqrt{-g}g^{xx}g^{11} A_1' \right )'- i\omega e^{-\phi} \sqrt{-g} g^{0x} g^{11} A_1'\\
&& - \left
( e^{-\phi}\sqrt{-g}g^{x0}g^{11} \left ( i \omega A_1+ikA_0 \right ) \right
)'\\
&& -e^{-\phi} \sqrt{-g} g^{00} g^{11} \left ( \omega^2 A_1 + \omega k A_0
\right )\end{array}
\end{equation}
The trick is to form a second order ODE for the gauge invariant combination
$E_1=k A_0+\omega A_1$.  We do this by putting the differential equations
into such form as the coefficients of $A_0''$ and $A_1''$ in the second and
third equations are unity then adding $k$ times the second equation to $\omega$
times the third equation.  We patch up the first derivative terms by adding
zero in the form given by (18).
\begin{equation}
\left ( i \omega A_0'+ik g^{11}g^{xx} A_1'+g^{11}g^{x0} \left (
k^2 A_0+\omega k A_1 \right ) \right ) \equiv 0
\end{equation}
We add this term with coefficient such that the first derivative terms add
up to a multiple of $k A_0+\omega A_1$.

The equation we finally obtain is
\begin{equation}
E_1''+ f_1 E_1' + f_2 E_1 =0
\end{equation}
Here one has
\begin{equation} \begin{array}{ccl}
f_1&=&\frac{5}{2x}- \frac{i \omega}{x^3 f}+ \\
&&\\
&&\frac{\frac{i k^2}{x^3}+ \omega \left (\frac{5}{2 x f}+
\frac{1}{2 x^4 f} \right )-\frac{2 i \omega^2}{x^3 f}+ \omega
\left (\frac{i \omega}{x^3 f}-\frac{5}{2 x} \right
)}{\omega-\frac{k^2}{\omega} f} \end{array}
\end{equation}
and
\begin{equation}\begin{array}{ccl}
f_2&=&-\frac{k^2}{x^3}+ \frac{i \omega}{2 x^4 f}+\frac{\omega^2}{x^3 f}+ \\
&&\\
&&\frac{k}{x^3} \left ( \frac{k \left ( \frac{i k^2}{x^3} + \omega
\left ( \frac{5}{2 x f}+ \frac{1}{2 x^4 f} \right ) -\frac{2 i
\omega^2}{x^3 f} \right )+ \omega \left ( \frac{i \omega
k}{x^3f}-\frac{5k}{2x} \right )}{i \left ( \omega^2 - k^2 f \right
)} \right )\end{array}
\end{equation}

After all this work, it is easy to check that this equation has the same $k \rightarrow 0$ limit as the transverse mode - the quasinormal frequencies are degenerate in this limit.  We also note that by rescaling $\omega \rightarrow \lambda^2 \omega$ and $k \rightarrow \lambda k$ and taking $\lambda \rightarrow 0$ one finds that a normalizable, regular solution to the longitudinal equation exists for $k=\omega=0$, which is just $E_1=x^{-\frac{3}{2}}$.  This additional mode is not present in any of the other spectra, and is related to the hydrodynamic behaviour of the field theory.

\begin{centering}
\begin{figure*}
\begin{centering} \begin{tabular}{cc}
\includegraphics[width=80mm]{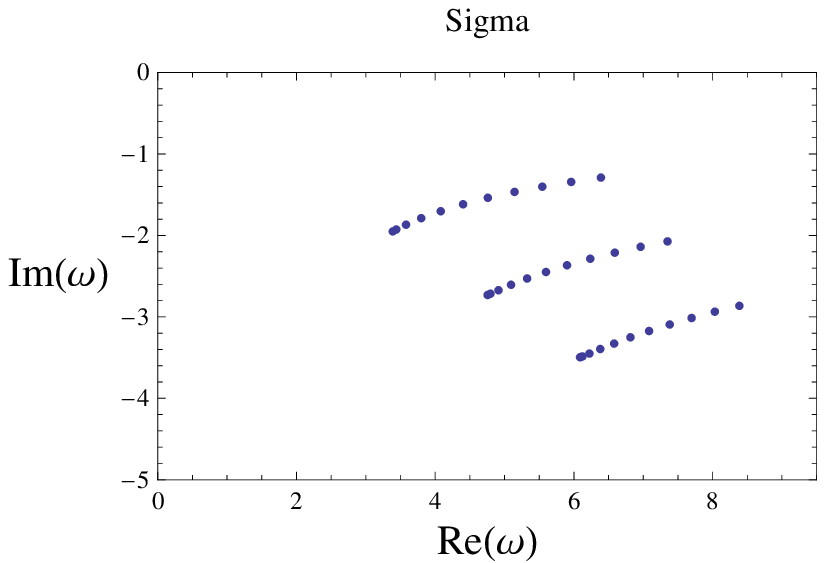}
\end{tabular}
\par
\end{centering}
\caption{\small{Lowest quasinormal frequencies at finite real spatial momentum
$k$ in the complex $\omega$ plane for the scalar
$\bar{q}q$ bound state.
The momentum $k$ ranges from $0$ to $5.0$ in steps of $0.5$ (in the same units as $\omega$) as one moves to the right in the plot.}}
\end{figure*}
\par
\end{centering}

\begin{centering}
\begin{figure*}
\begin{centering} \begin{tabular}{cc}
\includegraphics[width=80mm]{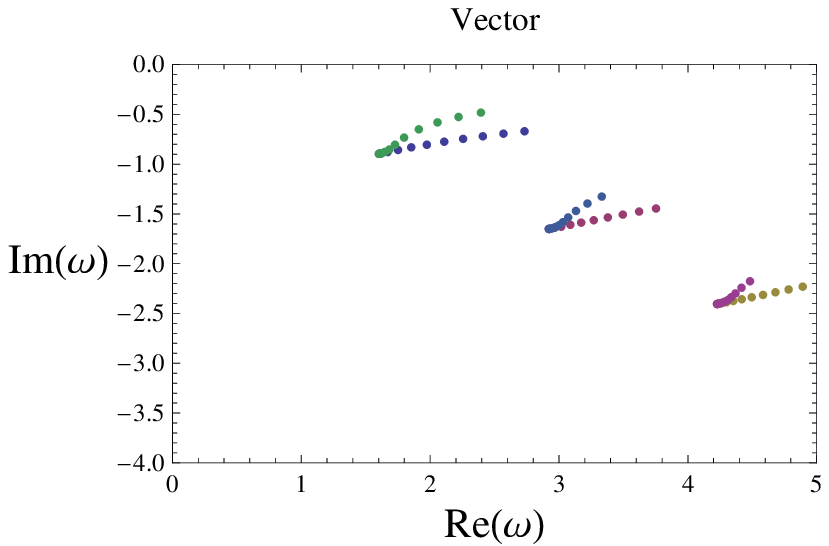} 
&
\includegraphics[width=80mm]{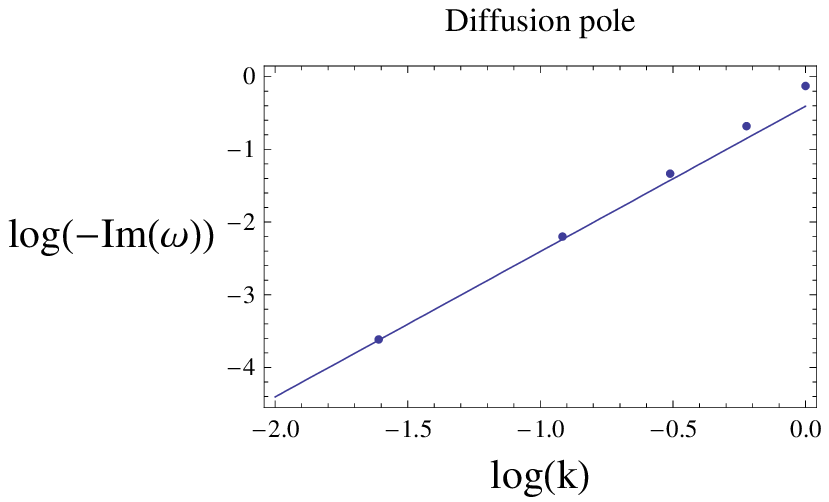}
\end{tabular}
\par
\end{centering}
\caption{\small{On the left we show the lowest vector $\bar{q}q$ bound state quasinormal frequencies
at finite real spatial momentum $k$ in the complex $\omega$ plane.
The momentum ranges from $0$ to $2.0$ in steps of $0.2$ as one moves to the right
(the lowest mode in the longitudinal spectrum, corresponding to the diffusion pole
in the hydrodynamic limit, is excluded).  For each mode the flatter trajectory corresponds
to the transverse species and the steeper to the longitudinal.  \\ \\ On the right we show
the momentum dependence of the lowest quasinormal mode for the longitudinal
vector meson.  The equation of the line is $y=\log \frac{2}{3}+2 x$ showing the validity of
the relation $\omega= - i D k^2$ for small $k$, with $D= \frac{2}{3}$.
}}
\end{figure*}
\par
\end{centering}

\section{Finite spatial momentum}

We can obtain an effective dispersion relation for our modes ie a function $\omega(k)$.
This is done by solving the wave equations obtained in the
previous sections for general complex $k$. Similar computations in the D3/D7 system can be
found in \cite{Ejaz:2007hg}.

Real momentum $k$ corresponds to a state which is a travelling
plane wave on the Minkowski spacetime of the dual field theory.
Switching on a finite real $k$ in the equation and using shooting
we have found the behaviour of the quasinormal frequencies in the
complex $\omega$ plane.  It is found that the states become more
massive and more stable as $k$ is increased.

The results for the first three quasinormal modes for the scalar
$\bar{q}q$ bound state are plotted in Fig.4.
In Fig.5 we show the first three vector quasinormal modes for the transverse
and longitudinal modes.
These are degenerate for $k=0$ but behave differently as $k$ is increased -
the main difference is that the longitudinal states become more stable but less massive relative
to the transverse states for the same $k$ as $k$ is increased.

Finally we note that, as mentioned above, there is an additional quasinormal mode for
the longitudinal electric field component of the Maxwell field on the flavour brane.
For small $k$ this lies close to the origin and on the imaginary axis.
As shown in for example \cite{Myers} the diffusion coefficient $D$ for
flavoured fundamental matter can be computed from this state.
For small spatial momentum $k$ it
obeys the `hydrodynamic' relation $\omega = -i D k^2$.
Here we test whether this relation can be obtained using our ingoing coordinates.
The calculation we do is then the generalization of the calculation done for the
vector meson, including a nonzero spatial momentum, and examining the gauge-invariant
longitudinal electric field component. In Fig.5 we plot the position of this pole
on the imaginary axis as a function of $k$. We extract $D=\frac{2}{3}$.

We note our result for the diffusion coefficient is related to the value obtained in \cite{Myers}
which is found to be $D=\frac{1}{2 \pi T}$.  We are measuring in units of$\sqrt{\frac{R^3}{u_T}}$
so we obtain the numerical value
\begin{equation}
D=\frac{1}{2 \pi} \frac{4 \pi}{3} \equiv \frac{2}{3}
\end{equation}
Our result therefore matches that of \cite{Myers} providing a check on our numerics
(albeit for small $\omega$ and $k$).

\section{Conclusion}

We have found the quasinormal frequencies for a variety of
different species (the Klein-Gordon scalar, scalar quark
bound states and vector mesons) in the Sakai-Sugimoto model at
high temperature. A crucial part of the analysis was to change
coordinates so that the infalling quasi-normal modes become
regular at the horizon so numerical shooting becomes
straightforward. It is noteworthy that in these coordinates the
equations for the  Klein-Gordon scalar, the scalar $\bar{q}q$ and
the transverse vector $\bar{q}q$ all have the form
\begin{equation} \begin{array}{ccc}
\left ( x^n f \phi' \right )' - i
\omega \left ( 2 x^{n-3} \phi'+(n-3) x^{n-4} \phi \right ) &&\\ &&\\
+\left (\omega^2-k^2 \right ) x^{n-3} \phi=0\end{array}
\end{equation}

For the Klein-Gordon scalar $n=\frac{19}{4}$, for the sigma $n=\frac{11}{2}$ and for the vector $n=\frac{5}{2}$.
This means the quasinormal spectra look extremely similar.  The only thing making the frequencies
different is the value of $n$.  The effect of increasing the value of $n$ is to move
the quasinormal modes out from the origin in the complex frequency plane.

We also computed these states at finite real momenta where the modes become more massive and more stable.
We have obtained the numerical value for the diffusion coefficient for fundamental flavoured
matter and our result is consistent with the calculation of \cite{Myers}.
\vspace{1cm}

\noindent {\bf Acknowledgements:}
ET would like to thank one of the authors of \cite{Dolan:2005a} (SD)
for helpful comments regarding the transformation to ingoing coordinates
and STFC for his studentship funding.

\end{document}